\documentstyle[11pt,psfig]{article}
\def\simge{\mathrel{%
   \rlap{\raise 0.511ex \hbox{$>$}}{\lower 0.511ex \hbox{$\sim$}}}}
\def\simle{\mathrel{
   \rlap{\raise 0.511ex \hbox{$<$}}{\lower 0.511ex \hbox{$\sim$}}}} 
\def\slashchar#1{\setbox0=\hbox{$#1$}           
   \dimen0=\wd0                                 
   \setbox1=\hbox{/} \dimen1=\wd1               
   \ifdim\dimen0>\dimen1                        
      \rlap{\hbox to \dimen0{\hfil/\hfil}}      
      #1                                        
   \else                                        
      \rlap{\hbox to \dimen1{\hfil$#1$\hfil}}   
      /                                         
   \fi}                                         %

\def\ts{\thinspace}

\def\ra{\rightarrow}

\def\ol{\bar}

\def\be{\begin{equation}} 
\def\ee{\end{equation}} 
\def\bea{\begin{eqnarray}}
\def\eea{\end{eqnarray}}
\def\ba{\begin{array}}
\def\ea{\end{array}}
\def\chipr{\chi^{\ts \prime}}

\def\Ntc{N_{TC}}

\def\kslash{\raise.15ex\hbox{/}\kern-.57em k}
\def\tro{\rho_{T}}

\def\troz{\rho_{T}^0}
\def\tom{\omega_T}
\def\tpi{\pi_T}
\def\tpipm{\pi_T^\pm}
\def\tpimp{\pi_T^\mp}
\def\tpip{\pi_T^+}
\def\tpim{\pi_T^-}
\def\tpiz{\pi_T^0}
\def\tpipr{\pi_T^{0 \prime}}

\def\toppip{\pi_t^+}

\def\toppiz{\pi_t^0}

\def\gev{{\rm GeV}}

\begin{document}
\title{Technihadron Production and Decay at LEP2}
\author{Stephen Mrenna\thanks{mrenna@physics.ucdavis.edu}\\
Department of Physics, University of California at Davis\\
\large Davis, CA  95616}
\maketitle

\begin{abstract}
The simple ``straw--man'' model of low--scale technicolor
contains light color--singlet technihadrons, which mix
with the electroweak gauge bosons.
We present lepton collider production rates at the parton level,
and show that experiments at {\sc LEP2} 
may be sensitive to the presence of technirho
and techniomega states with masses $10-20$ GeV {\it beyond} the 
center--of--mass energy because of the mixing.
The exact sensitivity depends on several parameters,
such as the technipion mass, the technipion mixing angle, and the
charge of the technifermions.
In an appendix, we describe the implementation of the model
into the event generator {\sc PYTHIA} for particle--level studies
at lepton and hadron colliders.
\end{abstract}

\vspace{-12cm}
\font\fortssbx=cmssbx10 scaled \magstep2
\hbox to \hsize{

$\vcenter{
\hbox{\fortssbx University of California - Davis}
}$

\hfill

$\vcenter{\normalsize
\hbox{\bf UCD-99-13} 
}$}


\newpage

\section{The Technicolor Straw Man Model}

Strongly--coupled models of electroweak symmetry breaking are expected
to have additional structure beyond the would--be Goldstone bosons that
give mass to the $W$ and $Z$ bosons.
In this study, we predict the lepton--collider production rates 
at the parton level
of the lightest color--singlet technivector mesons $\tro$ and
$\tom$ with masses around 200 GeV, which should be
relevant to physics studies at {\sc LEP2}. 
The basis for this analysis is the ``Technicolor
Straw Man Model,'' or {\sc TCSM}~\cite{tcsm}, which consists of a particle
spectrum and effective Lagrangian to 
describe the phenomenology of the lowest--lying states from
a more complete theory of dynamical symmetry breaking.
The complete theory is expected to contain some of
the aspects of technicolor~\cite{tc}, extended technicolor~\cite{etc},
walking technicolor~\cite{wtc}, top
condensate models and topcolor-assisted technicolor
({\sc TC2})~\cite{topcondref,topcref,tctwohill,tctwoklee}, and/or multiscale
technicolor~\cite{multi}.  Some signatures of low--scale technicolor in
the {\sc TCSM} have been considered at hadron
and muon colliders~\cite{elw}.
Here, we address the issue of what can be learned from the LEP2 collider
operating at a center--of--mass energy $\sqrt{s}\simeq 200$ GeV.
We concentrate on the challenging case when
the $\tro$ and $\tom$ masses are larger than $\sqrt{s}$.

In the {\sc TCSM}, only
the lowest-lying bound states of the lightest technifermion doublet, $(T_U,
T_D)$ are considered.  
The technifermions are  assumed to be color singlets and to 
transform under technicolor $SU(\Ntc)$ in a fundamental representation,
with electric charges $Q_U$ and $Q_D$.  The phenomenology considered
here depends only on the sum of these charges $Q\equiv Q_U+Q_D$.
The bound states of the technifermions  are the pseudoscalar isotriplet
$\Pi_T^{\pm,0}$ and isosinglet $\Pi_T^{0 \prime}$ mesons, 
and the vector isotriplet $\tro^{\pm,0}$ and isosinglet
$\tom$ mesons. 
The technihadron mass scale is set by the technipion decay constant $F_T$.
In TC2 models, $F_T \simeq F_\pi/\sqrt{N_D}$, where $F_\pi=246\,\gev$, and $N_D$ 
is the the number of electroweak
doublets of technifermions.  In a specific model,
$N_D \simeq 10$ and $F_T \simeq 80\,\gev$ \cite{tctwoklee}.
The interaction states $\Pi_T$ are admixtures of the electroweak Goldstone bosons $W_L$ and
the  mass eigenstates of
pseudo--Goldstone technipions $\tpipm, \tpiz$:
\be\label{eq:pistates}
 \vert\Pi_T\rangle = \sin\chi \ts \vert
W_L\rangle + \cos\chi \ts \vert\tpi\rangle\ts,
\ee
where $\sin\chi = F_T/F_\pi$ ($\simeq 1/\sqrt{10}$ in the
model mentioned above). Similarly, $\vert\Pi_T^{0 \prime} \rangle
= \cos\chipr \ts \vert\tpipr\rangle\ + \cdots$, where $\chi'$ is another
mixing angle and the ellipsis refer to other technipions needed to eliminate
the technicolor anomaly from the $\Pi_T^{0 \prime}$ chiral current. 
If techni--isospin is a good approximate symmetry,
$\tro$ and $\tom$, and, separately, $\tpiz,\tpipr,\tpipm$  are
nearly degenerate in mass.  However, 
there may be appreciable $\tpiz$--$\tpipr$
mixing \cite{elw}. If that is the case,
the lightest neutral technipions are maximally--mixed
$\ol T_U T_U$ and $\ol T_D T_D$ bound states.

\subsection{Techniscalar decays}

\begin{figure}
\psfig{file=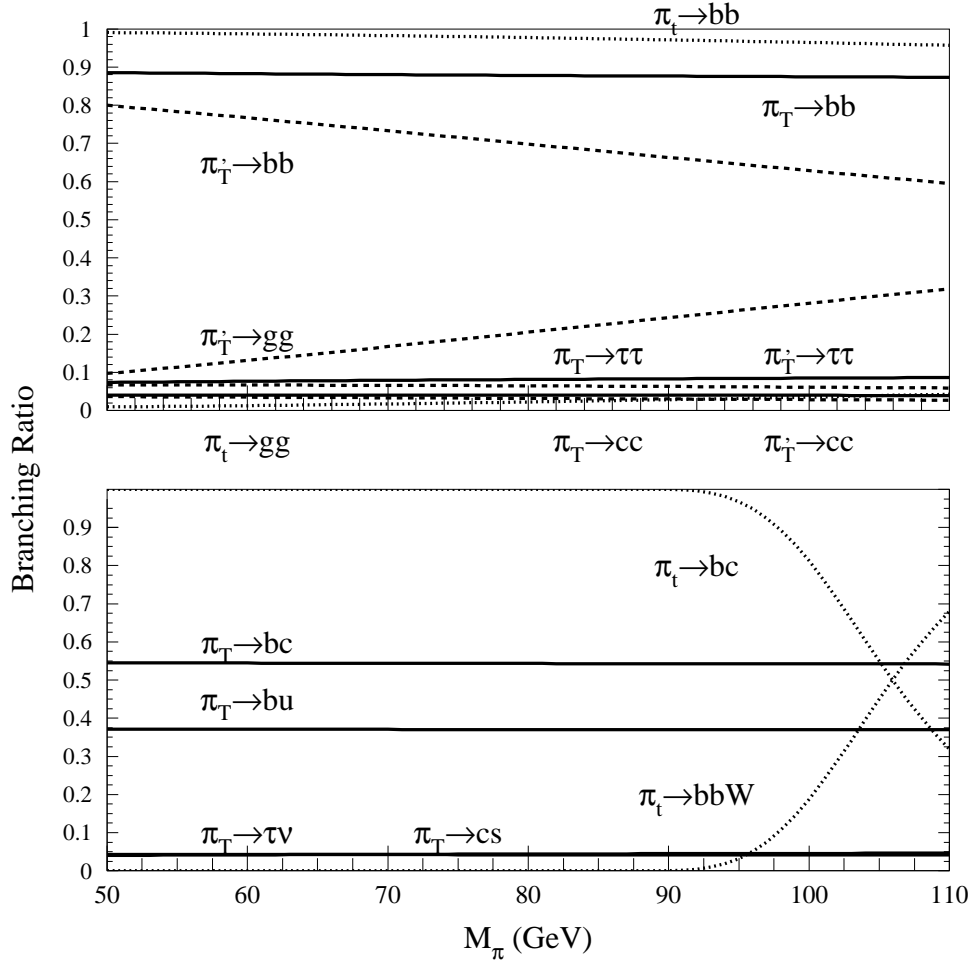,width=14cm}
\caption[]
{Comparison of different technipion and toppion branching
ratios.  Decays of $\tpiz$, $\tpipr$, and $\pi_t^0$ are illustrated
in the upper part, while $\pi_T^\pm$ and $\pi_t^\pm$ are shown in
the lower part.  The model assumptions are described in the text.}
\label{brpi}
\end{figure}

Technipion decays are induced mainly by extended technicolor (ETC)
interactions which couple them to quarks and leptons like Higgs bosons~\cite{etc}. 
With a few exceptions, technipions are expected to decay into the
heaviest fermion pairs allowed. One exception is that decays to top quarks
are not enhanced, since ETC interactions only generate a few GeV of the top 
quark mass.  Another exception is that the constituents of the
isosinglet $\tpipr$ may include colored technifermions as well as
color-singlets, so that decays into a pair of gluons are possible.
Therefore, the important decay modes are $\tpip \ra c \ol b$,$u\bar b$,
$c \ol s$, $c\bar d$ and $\tau^+ \nu_\tau$;
 $\tpiz \ra b \ol b$, $c\ol c$, $\tau^+\tau^-$; and 
$\tpipr \ra gg$, $b \ol b$, $c \ol c$,
$\tau^+\tau^-$. Branching ratios are presented in Fig.~\ref{brpi} 
for $\tpiz$ (solid
lines) and $\tpipr$ (dash--dot lines) using the
expressions of Ref.~\cite{tcsm} and $C_f=1$, except $C_t=m_b/m_t$, $C_{\pi_T}=4/3$,
$N_{TC}=4$, and $F_T=82$ GeV.  The $\tpiz$ and $\tpipm$ branching ratios are 
fairly flat as a function of $M_{\tpi}$, while $\tpipr$ shows more
variation because of the $gg$ decay mode.

In addition to these considerations, there may be light topcolor pions
present in a realistic theory, and these can mix with the ordinary
technipions.  The topcolor pions couple preferentially to top quarks,
but there can be flavor mixing and instanton effects \cite{burdman}.  
The neutral
top pion $\toppiz$ can decay $\to t\bar t$ above threshold;
$\to t\bar t^*\to tbW$ below threshold; $\to t\bar c,t\bar u$ through mixing;
$\to b\bar b$ through instanton effects; or $\to gg$ through a top quark loop.
The charged top pion can decay
$\toppip\to t\bar b$ above threshold; 
$\to t^*\bar b\to b\bar bW$ below threshold;
or $\to b\bar c$ (etc.) through mixing.    
Typical branching ratios for $\pi_t^0$ and $\pi_t^\pm$ decays
are shown in Fig.~\ref{brpi} (short--dashed lines) with the
toppion decay constant set to 82 GeV.  For the mass range
considered here, only $\pi_t^0$ decays to $b\bar b$ and $gg$ final states
are important.  Note that off--shell decays
$\pi_t^\pm\to b\bar bW$ can be competitive with the mixing--suppressed decay
to $bc$ (the suppression was arbitrarily chosen as $(.05)^2$ for this
plot).  
In the following, we ignore the complication of technipion--toppion mixing and
assume that the technipions decay according to the
expectations of the {\sc TCSM}.

\subsection{Technivector decays}

In the limit that the electroweak gauge couplings $g,g'\to 0$, 
the isospin--conserving decays of $\tro$ and $\tom$ are fixed by
the technipion mixing angle:
\bea\label{eq:vt_decays}
\tro &\ra& \Pi_T \Pi_T = \cos^2 \chi\ts (\tpi\tpi) + 2\sin\chi\ts\cos\chi
\ts (W_L\tpi) + \sin^2 \chi \ts (W_L W_L) \ts; \nonumber \\
\tom &\ra& \Pi_T \Pi_T \Pi_T = \cos^3 \chi \ts (\tpi\tpi\tpi) + \cdots \ts.
\eea
Because of the lifting of the technipion masses by the hard
technifermion masses, the {\sc TCSM} assumes the decay $\tom\to\tpi\tpi\tpi$
are kinematically forbidden.  In addition, we do not consider models
where $\tom\to W_LW_LZ_L$ is possible.
The isospin violating decay rates obey the relation
$\Gamma(\tom \ra \pi^+_A \pi^-_B) = \vert\epsilon_{\rho\omega}\vert^2 \ts
\Gamma(\troz \ra \pi^+_A \pi^-_B)$,
where $\epsilon_{\rho\omega}$ is the isospin-violating $\tro$-$\tom$ mixing
amplitude.  In QCD, $\vert \epsilon_{\rho\omega}\vert \simeq 5\%$, so
the isospin violating decays in the {\sc TCSM} are expected to be unimportant.

The technivectors also undergo 2--body decays to transverse gauge bosons
and technipions ($\gamma\tpi$, $W\tpi$, {\it etc.}) 
and fermion--anti-fermion pairs $f\bar f$.
The decay rates to transverse gauge bosons are set by a vector or axial mass
parameter, $M_V$ and $M_A$, respectively,
which is expected to be of the same order as $F_T$,
and are proportional to $\cos^2\chi$ or $\cos^2\chi'$.
Decays where the mother
and daughter techniparticle have the same isospin and electric charge
are proportional to $Q^2$, and the decays to $Z^0\pi_T$ are of similar strength
as $\gamma\pi_T$.
The $\tro$ and $\tom$ decay to fermions because of the
technifermion couplings to the standard model (SM) gauge bosons.  In
general, the branching ratios to fermions are small, and
the $\tom$ decay rate is proportional to $Q^2$.

\subsection{Direct technipion production}
The lightest technimeson states are difficult
to produce directly at $e^+e^-$ colliders.  
The process $e^+e^-\to \pi_T^0
\propto \Gamma(\pi_T^0\to e^+e^-) \propto ({m_e\over F_T})^2$ is
suppressed by a small coupling, while
$\gamma\gamma\to \pi_T^0
\propto \Gamma(\pi_T^0\to \gamma\gamma)$ is one--loop
suppressed.  Additionally, the technipions have no tree level couplings to
$W$ or $Z$, negating the usual Higgs boson production
modes at lepton colliders.  The charged technipion can
be pair--produced through a virtual photon, but the
production rates are not large.  For a center--of--mass
energy $\sqrt{s}=200$ GeV, the production cross section
falls as (.169,.115,.063,.024,.011) pb for $M_{\pi_T^\pm}=$
(80,85,90,95,97) GeV.  The SM $W^+W^-$ cross section is
about 20 pb, and it is problematic whether an excess of
events with heavy flavor can be observed above
the backgrounds (because of {\sc TC2}, such a light
charged technipion is not constrained by top quark decays).
Presently, LEP experiments set a 95\% C.L. exclusion on
a charged Higgs boson with mass in the 
range $52-58$ GeV \cite{aleph_chargedhiggs}.  Therefore, we only
consider models with technipions heavier than this limit.

\subsection{Technivector production}

The explicit formulae for the cross sections of the technivector--mediated
processes have been presented in Ref.~\cite{tcsm}.  
Unlike the technipions, the technirho and techniomega can have
substantial mixing with the SM gauge bosons, and can be
produced with electroweak strength.  The mixing between
$\gamma,Z$ and $\tro,\tom$ is proportional to $\sqrt{\alpha/\alpha_{\tro}}$,
where $\alpha$ is the fine structure constant and $\alpha_{\tro}$ is
the technirho coupling, which is fixed in the {\sc TCSM} by scaling
the ordinary rho coupling by $N_{TC}$ ($=4$ in this analysis).  
The full expression for the mixing depends also on the masses and
widths of the technivectors.  In addition, $\gamma-\tom$ and $Z-\tom$
mixing is proportional to $Q$.

From the discussion of decay rates above, the technirho and
techniomega are also expected to be narrow, which naively reduces
the reach of a lepton collider to regions where the center--of--mass energy is
close to the resonance mass.  However, the resonances are not 
of the simple, Breit--Wigner form, and the effects of mixing
can be seen at lower energies than a few total widths from the
resonance mass.  On resonance, the production cross
sections are substantial (${\cal O}$(nb) strength for the models considered here),
and a tail may be visible even if the nominal mass of the resonance
is $10-20$ GeV {\it above} the collider energy.
If the resonance mass is substantially {\it below}
the center--of--mass energy, then the resonance is produced in radiative
return events, and should be easily excluded~\cite{landsberg}.

\section{Technivector models}

To estimate the prospects for observing technivectors at LEP2,
we have to fix all of the {\sc TCSM} parameters.  The remaining 
important parameters
are the mass splittings
between the vectors and scalars $\Delta M\equiv M_{\tro}-M_{\tpi}$,
the technipion mixing angle $\chi$,
and the sum technifermion charge $Q$.
The choices of model parameters are outlined in Table 1.  The
vector and axial mass parameters are
fixed at $M_V=M_A=100$ GeV, 
$\sin\chi'=\sin\chi$, and $M_\rho=M_\omega$ for simplicity.
While the choice is not exhaustive, the models are intended to
illustrate basic patterns of signals.

\begin{table}
\begin{tabular}{||c||c|c|c|c||c||c||}
\hline
~~Model~~ & ~~$\sin\chi$~~  & ~~$Q$~~  & ~~$M_\rho$ (GeV)~~  &  ~~$M_\pi$ (GeV)~~ & $\Gamma_{\tro}$ ($\Gamma_{\tom}$) (GeV) & Symbol \\ \hline\hline
1 & 1/3 & 5/3 & 210 & 110 & 1.36 (.46) & dashes \\
2 & 1/3 & 0 & 200 & 110 & .33 (.34$\times 10^{-2}$) & dots    \\
3 & 1/3 & $-1$  & 205 & 175 & .15 (.26$\times 10^{-1}$) & dash-dots  \\
4 &   1 & 5/3 & 200 & 105 & 7.64 (.44$\times 10^{-1}$) & $\triangle$ \\
5 &   0 & 5/3 & 200 & 105 & 0.64 (.43) & $+$ \\
6 &   0 & 5/3 & 205 & 100 & 1.24 (.56) & $\circ$ \\
7 &   0 & 0 & 200 & 80 & 8.50 (0.23) & $\times$   \\
8 &   1 & 0 & 200 & 80 & 7.67 (0) & $\Box$  \\ \hline
\end{tabular}
\caption{The parameters of the {\sc TCSM} models used in this analysis.}
\end{table}

Model 1 has relatively heavy $\tro$ and $\tom$, and
the decays $\tro\to W_L^+W_L^-$ and $\to W_L^\pm\tpimp$ are suppressed by mixing
and phase space.  The charge $Q$ is large, so that
$\tom$ has a large branching ratio to $\gamma\pi_T$ and
$f\bar f$ final states and a large $\gamma-\tom$ and $Z-\tom$ mixing.
Model 2 has a lighter $\tro$ and $\tom$ and $Q=0$,
so that the $\tom f\bar f$ coupling and $\gamma/Z-\tom$ mixing vanishes.
Model 3 has a small mass splitting $\Delta M$, so that
$\tro\to W_L^\pm\tpimp$ is forbidden on--shell, and $Q=-1$, which yields
similar $\gamma/Z-\tro$ and $\gamma/Z-\tom$ mixing.
Model 4 has the maximal coupling to $W_L^+W_L^-$, while
Model 5 has the minimal coupling, but $\tro\to\tpip\tpim$
is kinematically forbidden on--shell.
Model 6 is similar to Model 5, but $\tro\to\tpip\tpim$ is allowed on--shell.
Finally, Models 7 and 8 have light technipions, with unsuppressed couplings
to $W_L^+W_L^-$ and $\tpip\tpim$, respectively, but $Q=0$ to suppress
$\tom f\bar f$ couplings and $\gamma/Z-\tom$ mixing.
The decay widths for the $\tro$ ($\tom$) 
calculated from these parameters are shown in the next--to--last 
column of Table 1.  The final column shows the symbols used
in the figures to denote the Models 1--8.

\section{Signatures}

We concentrate on four basic signatures.  The first two,
the Drell--Yan production of $\mu^+\mu^-$ and $W_L^+W_L^-$
pair production, contain only SM particles in the final state.
The last two contain either two technipions or a technipion
and an electroweak gauge boson.

\subsection{Drell--Yan}

\begin{figure}
\psfig{file=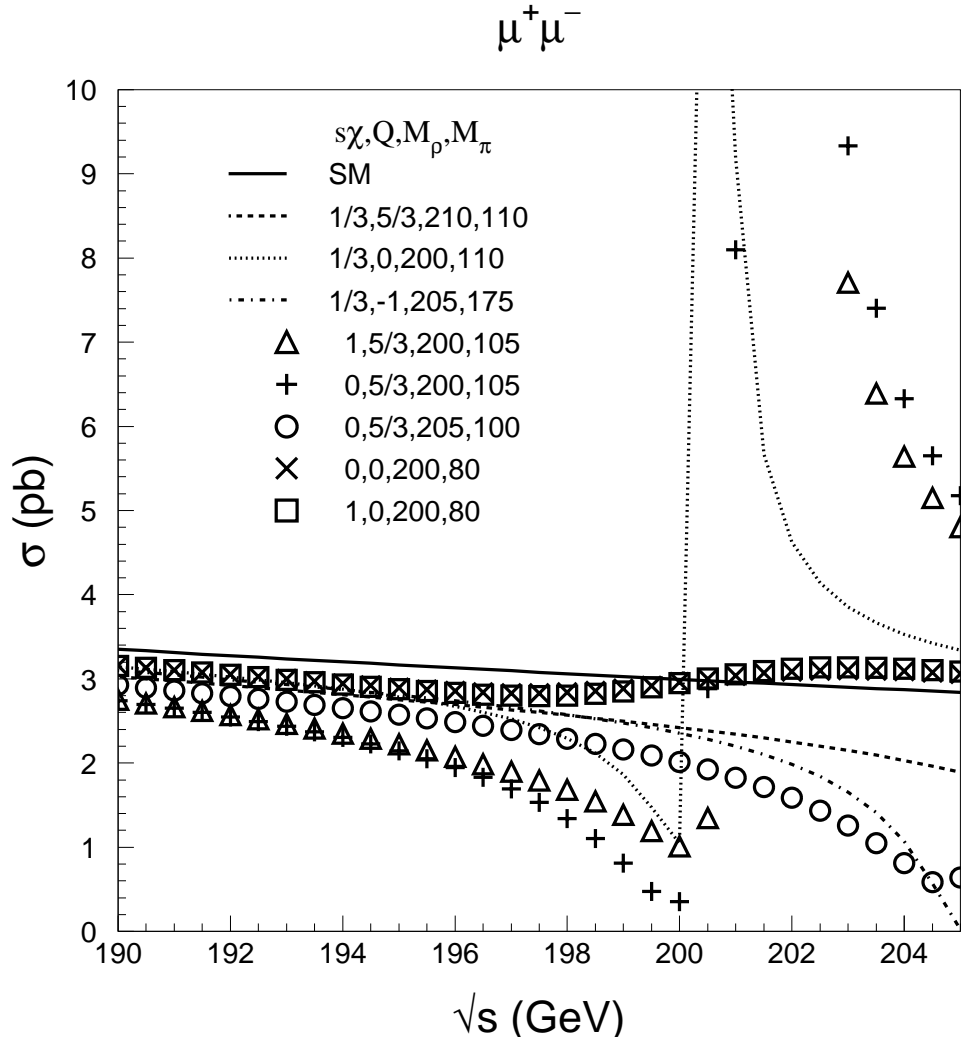,width=14cm}
\caption[]
{Signatures of {\sc TCSM} in the final state $\mu^+\mu^-$ at an $e^+e^-$ collider
with center of mass energy $\sqrt{s}$.  The standard model prediction
is shown as the solid (straight) line.}
\label{epem}
\end{figure}

As explained above, the technirho and techniomega couple to
final states containing fermion pairs through mixing with $\gamma$
and $Z$ bosons.  We consider the
the $\mu^+\mu^-$ final state here, since this avoids the complication
of Bhabha scattering.  The expected cross sections for the various
models as a function of the center-of-mass energy $\sqrt{s}$ is illustrated
in Fig.~\ref{epem}.  It is worth noting the sensitivity to resonances with pole
masses above 
the energy $\sqrt{s}$, despite the fact that the resonances
are narrow.  In particular, Model 1, 
with $M_{\tro}=210$ GeV and $\Gamma_{\tro}=1.36$ GeV,\footnote{The input mass parameters for the technivectors
are not pole masses, so the peak of the resonance is shifted.}
has $S/B=.03,.06,.19$ at  $\sqrt{s}=160,180,200$ GeV in
the $\mu^+\mu^-$ final state, where $S\equiv\Delta\sigma{\cal L}$ is 
the deviation from the standard model cross section times the integrated
luminosity, and $B$ is the expected number of standard model events.
The large values for $\Delta\sigma$ (even sizeable 50 GeV from the resonance peak)
is due to the large charge $Q=5/3$ in this model.
If $Q=-1$ ($Q_U=0$), then $S/B=.02,.04,.12$ with all other
parameters fixed.  Likewise, for $Q=0$,  which
is the limit that the $\tom f\bar f$ coupling and $\gamma/Z-\tom$ mixing vanishes,
we have $S/B=.01,.02,.07$.
Far from the resonance peak, measurements of such variations in the 
overall rate will have important systematic as well as statistical errors,
so it is important to have a verification of an effect.
Because of the SM quantum numbers (and the energy range considered),
there is more sensitivity in the lepton pair final state than
in the quark pair, and,
for the same set of parameters, the effect
in the $b\bar b$ final state is roughly half of that in the $\mu^+\mu^-$ one.
There is a also difference in the angular distribution of the decay
products because of the interference between the various resonances,
but this is not dramatic.  

The general feature that the cross section decreases before
increasing on the resonance is true even for $Q<0$, since
the $\gamma\gamma,ZZ$ and $\gamma Z$ components of the inverse
propagator are quadratic in $Q$.  The only models that do not
demonstrate a significant effect in the fermion pair final state
are those where the technirho is fairly wide
and $Q=0$, so that the $\gamma/Z-\tom$ mixing
vanishes (Models 7 and 8).\footnote{In this extreme case, the techniomega
appears to be unreasonably narrow.  Small  isospin--violating
effects will have to be included, but
they will not contribute significantly to the $f\bar f$ final state.}  In these
cases, a substantial signal is expected in the 
$\tro$--mediated $W_L^+W_L^-$ or $\tpip\tpim$ channels.

\subsection{$W_L^+W_L^-$}
\begin{figure}
\psfig{file=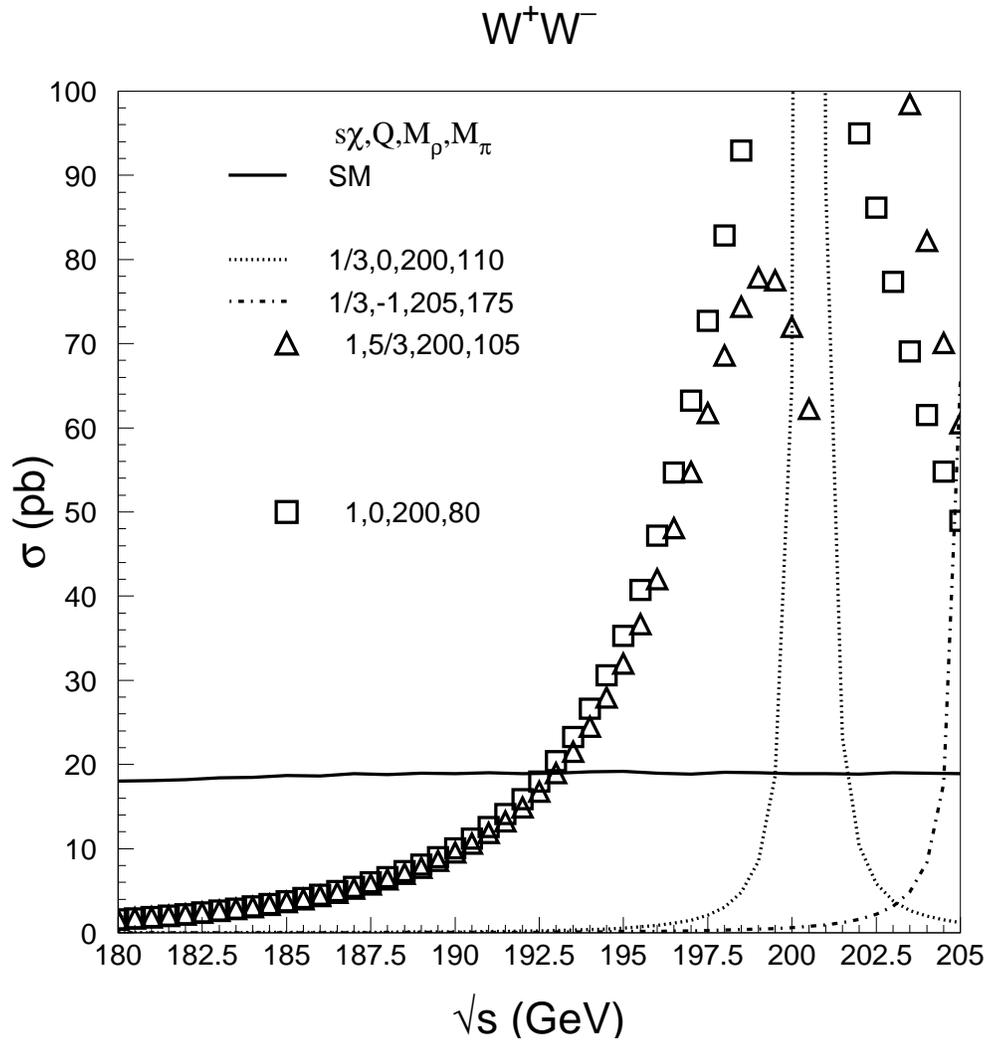,width=14cm}
\caption{Same as Fig.~\ref{epem}, except for the $W^+W^-$ final state.
The {\sc TCSM} contribution should be added to the standard model prediction
shown.}
\label{wpwm}
\end{figure}

If $\sin\chi\to 1$, the $\tro$ coupling to the $W_L^+W_L^-$ final state can
be important.  This is illustrated in Fig~\ref{wpwm}, where
only models that yield a visible signal are shown.  The SM prediction
for the $W^+W^-$ cross section is shown for reference.
Model 4 , with $\sin\chi=1$,
has $S/B=.09, 0.5, 3.8$ at $\sqrt{s}=180, 190, 200$ GeV, 
and Model 8 has a similar behavior.
(Note, in this figure, the {\sc TCSM} signal should
be added to the standard model component.)  
The technirho is fairly wide once the $W_L^+W_L^-$ channel is unsuppressed, but there is no
theoretical motivation for $\sin\chi\to 1$.  On the other hand, there is no realistic
theory, yet, so we present these results for completeness.
The feature around
$\sqrt{s}=200$ GeV in Model 4 arises from complicated $\gamma/Z-\tro$ interference.
When $\sin\chi=1/3$, the increase in cross section is limited to
a region of several GeV around the peak position, since the technirho is
much narrower.
Clearly, on or near the peak, the effect is a striking increase
in the total $W^+W^-$ production cross section.  Otherwise, the
signature is a moderate excess of $W_L^+W_L^-$ events 
on a potentially large background.

\subsection{$W_L^\pm\tpi^\mp+\tpip\tpim$}
\begin{figure}
\psfig{file=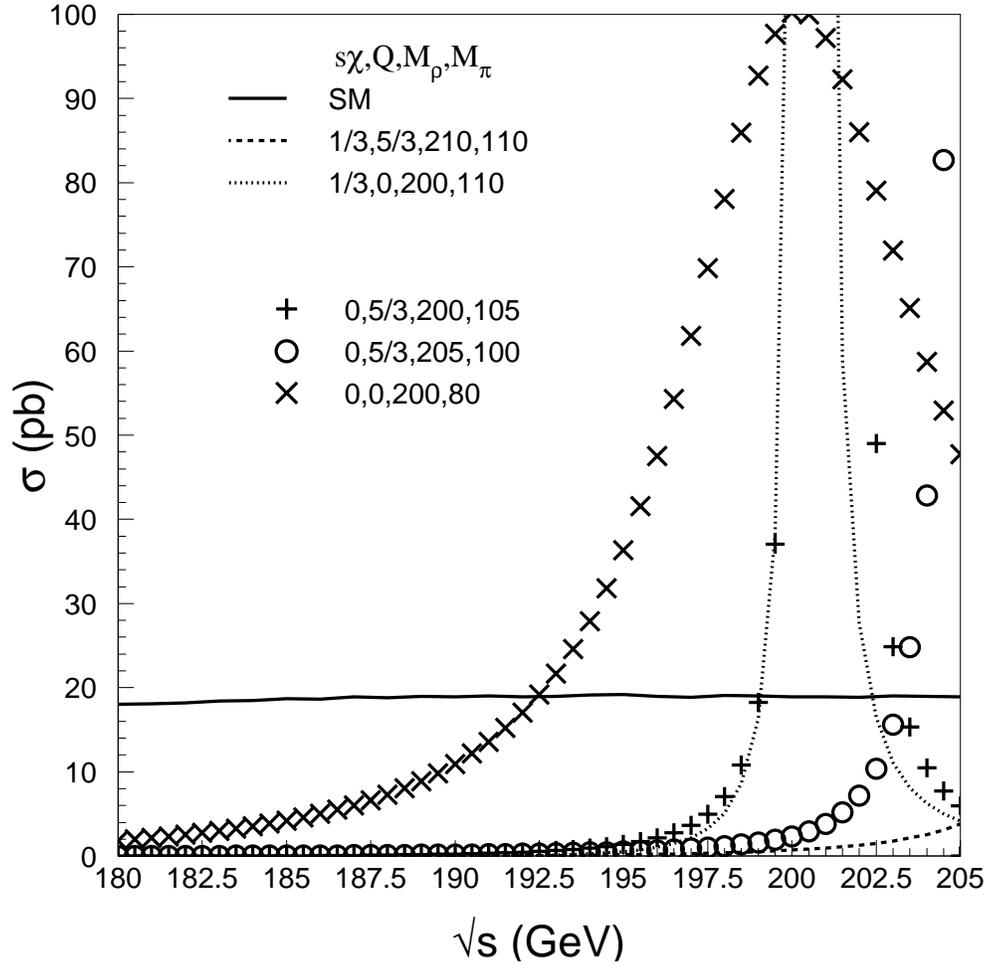,width=14cm}
\caption{Same as Fig.~\ref{wpwm}, except for the final state $W^\pm\pi_T^\mp$ 
and $\pi_T^\pm\pi_T^\mp$.  The standard model prediction
for $W^+W^-$ is shown as the solid line.  The {\sc TCSM} contribution 
prefers final states containing heavy flavor.}
\label{wtp}
\end{figure}

If the technipion is light enough, the $\tro$ coupling  to $W_L^\pm\tpimp$ and
$\tpip\tpim$ 
as $\sin\chi\to 0$
is complementary to the $W_L^+W_L^-$ coupling when $\sin\chi\to 1$.  This
is illustrated by comparing Model 7 in Fig.~\ref{wtp} to Model 8 in Fig.~\ref{wpwm}, 
which have large signals in one or the other channel.
Both models yield the same $S/B$ at $\sqrt{s}=180$ GeV in their respective channels.
Technirho and techniomega couplings to a transverse $W$ boson and $\tpipm$ also arise, but
typically at reduced rates compared to $W_L^\pm\tpimp$.
Since $\tpi^\pm$ decays preferentially to heavy flavor,
$W_L^\pm\tpimp$ or $\tpi^+\tpi^-$ production will produce an excess of
$\tau$ or
$b$ and $c$--tagged events in the total $W^+W^-$ data sample.  
The experimental sensitivity will be better if $M_{\tpi}$ is sufficiently
different from $M_W$.
Note that the off--resonance production
rate for $\tpi^+\tpi^-$ is generally much larger than the usual
charged Higgs boson pair production rate discussed earlier.

\subsection{$\gamma\tpiz,\gamma\tpipr$}
\begin{figure}
\psfig{file=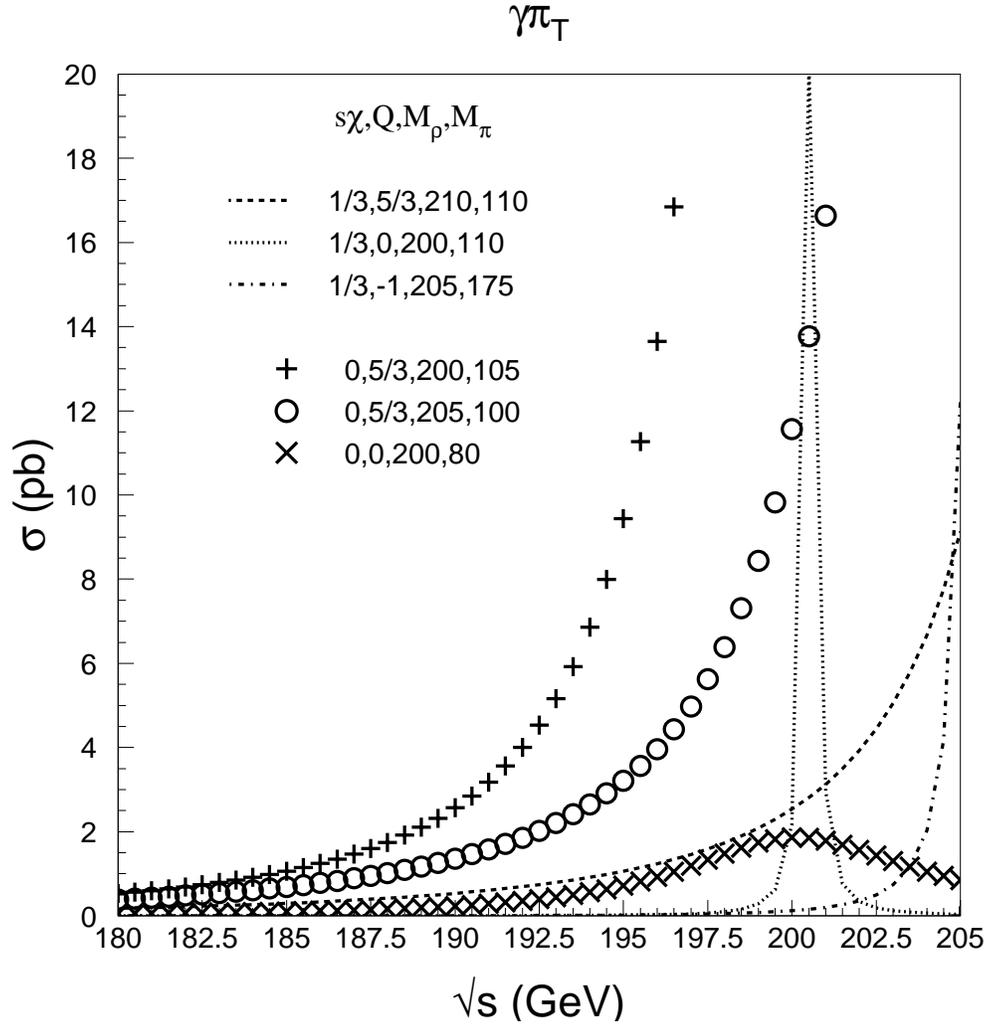,width=14cm}
\caption{Same as Fig.~\ref{wpwm}, except for the final state $\gamma\pi_T^0$ and
$\gamma{\pi_T^0}^{'}$.
The $\tpi$ decays
preferentially to heavy flavor or $gg$.}
\label{gampi}
\end{figure}

For $\sin\chi \simeq 0$, 
a significant $\gamma\tpiz,\gamma\tpipr$ signature can arise.
$Z\tpi$ production, while possible, is never 
important relative to $\gamma\tpi$ from phase space considerations.
The $\tom$ contribution to $\gamma\tpi$ can be enhanced significantly if $Q$ is large,
since the $\gamma/Z-\tom$ mixing is proportional to $Q$.
The expected
signal rate is shown in Fig.~\ref{gampi} for the various models.  We have
not attempted to estimate the backgrounds, which may be prodigious if
$M_{\tpi}\simeq M_Z$.
However, if $M_\pi$ is sufficiently different
from $M_Z$,
an off--resonance signal may be observable.
The expected final states are $\gamma b\bar b$, 
$\gamma\tau\tau$ or $\gamma gg$.  
On resonance, the $\gamma\tpi$ production rate 
can be ${\cal O}$(100 pb) or larger, and there is still some
rate off resonance even when the $\tro$ and $\tom$ are narrow.
Model 1 (with $Q=5/3$) yields a raw event rate of $.18, .54, 2.5$ pb
at $\sqrt{s}=180, 190, 200$ GeV.  This drops to $.06, .18, .90$ pb if $Q=-1$, and
$.01, .03, .15$ pb for $Q=0$.  These three choices for $Q$ represent
$\tom$ domination, equal $\tro$ and $\tom$ contributions, and $\tro$ domination.
Model 5, which has $\sin\chi=0$ and lighter
$\tro$ and $\tom$, has a raw event rate of $.53, 2.6, 271$ pb.

\section{Discussion and Conclusions}
We have presented examples of how several models of low--scale
technicolor, in the framework of the {\sc TCSM}, would manifest themselves
at a lepton collider operating near $\sqrt{s}=200$ GeV.  These
can be used to guide searches
at {\sc LEP2} to discover or constrain
{\sc TCSM} models.  The actual limits will depend on the
collider energy, the amount of delivered luminosity, and
the SM backgrounds in each channel.  For reference, it is
quite possible that LEP2 will operate at $\sqrt{s}=200$ GeV,
with 200 pb$^{-1}$ of data delivered to each experiment.
In this case, each experiment will be sensitive to cross sections
near 15 fb in channels which are relatively background free.
The production rates shown have no event selection cuts and no effects of
initial state radiation.  A dedicated analysis at the 
particle--level is now
under way \cite{kunin} based on the {\sc PYTHIA} event
generator~\cite{pythia}.  The details of how to
study the {\sc TCSM} using {\sc PYTHIA} are included
in the Appendix.  Here, we review the results of our 
parton--level study.

On or near resonance, there are substantial
signals of technirho and techniomega production
in one or more final states.  The typical width of the
$\tro$ considered is a few tenths to a few GeV, while the $\tom$
ranged from exceedingly narrow up to a few tenths of a GeV.
When $\sin\chi\to 1$, the decays $\tro\to W_L^+W_L^-$ are unsuppressed.
Likewise, when $\sin\chi\to 0$, but $\tro\to\tpip\tpim$ is kinematically
allowed, a complementary signature arises in 
the $\tpip\tpim$ final state, where $\tpipm$ decays predominantly
to heavy flavor.  For intermediate values of $\sin\chi$, decays
to $W_L^\pm\tpimp$ will occur when kinematically allowed. 
Also, there can be signals
in $f\bar f$ or $\gamma\tpi$ final states.  
These signatures should be unmistakable, since the on--resonance
cross sections can be of ${\cal O}$(nb).  For the same reason, we 
expect that technivectors with mass significantly below the center--of--mass
energy can be easily excluded by searching for radiative return events,
but this requires a detailed study \cite{landsberg}.

Because of the mixing between the technivector mesons and the
electroweak gauge bosons, signatures are not limited to
be near the resonance peak.  In particular, the presence of
the $\tro$ or $\tom$ may be inferred from 
a significant decrease in the $\mu^+\mu^-$ rate,
unless the $\gamma/Z-\tom$ mixing is small or the $\tro$ has
a width of several GeV.
The $b\bar b$ final state would yield a similar
effect, but at only about 1/2 the magnitude.   Also, the $\tom$ 
and $\tro$
can mediate the $\gamma\tpi$ final state, which may be observable
above backgrounds, provided that $M_{\tpi}$ is far enough from $M_Z$.  
Event rates of .18 pb are 
possible at 30 GeV below the resonance peak in the models
considered here,
depending on the technipion mass,
the technifermion charge $Q$, and the technipion mixing angle $\chi$.
Even rates closer to the resonance are much larger.

The choices of {\sc TCSM} parameters used in this analysis
were motivated by the beam energy of LEP2.  However,  
several technicolor--motivated analyses
have emerged based on the Run I data sets at the Tevatron~\cite{runi_searches}
that constrain the properties of the color--singlet technirho and techniomega.
In general, the technivector masses of the models considered here
are beyond
the sensitivity of these analyses, except for the techniomega
search, which may exclude the models with large $Q=5/3$ at
the 90\% C.L.  Therefore, it is expected that LEP2 can set
stronger limits than Run I at the Tevatron for $\tro^0$
and $\tom^0$ signatures for certain choices of {\sc TCSM} parameters.

In conclusion, unless the technipion masses are fairly light
compared to the technivector masses (which is not expected due
to the enhancement of the hard technifermion masses), or the
technipion mixing angle
$\sin\chi\to 1$ (which is not expected due to the large number
of technifermion doublets required in a model with a running
coupling), technivector--mediated $\mu^+\mu^-$ and $\gamma\tpiz,\gamma\tpipr$ final
states can be studied at LEP2 to discover or constrain simple models of
technicolor at collider energies substantially below the technivector masses.
The actual limit will depend on a detailed background analysis, but the
models studied here yield substantial effects at $10-20$ GeV below $M_{\tro}=M_{\tom}$.
The technirho alone can still produce visible
effects in these channels, or (1) the $W_L^\pm\tpi^\mp$ final states,
if kinematically allowed, (2) the $W_L^+W_L^-$ final state, if $\sin\chi\to 1$, 
or (3) the $\tpip\tpim$ final state, if $\sin\chi\simeq 0$ and the technipion
is light.

\section*{Acknowledgements}
I thank A.~Kounine, G.~Landsberg, and K.~Lane for useful
comments.  This work was inspired by the ``Strong Dynamics
for Run II Workshop'' at Fermilab.

\vfil\eject

\section*{Appendix}

The simulation of the production and decays of technicolor particles
has been substantially upgraded in 
{\sc Pythia v6.126}, which is available at moose.ucdavis.edu/mrenna, along
with documentation.

The full set of processes are:
\begin{verbatim}
* Drell--Yan (ETC == Extended TechniColor)                                    
194   f+fbar -> f'+fbar' (ETC)      
195   f+fbar' -> f"+fbar"' (ETC)
\end{verbatim}

The final state fermions are
$e^+e^-$ and $e^\pm\nu_e$, respectively, which can be changed through the
parameters {\tt KFPR(194,1)} and {\tt KFPR(195,1)}, respectively.

\begin{verbatim}
* techni_rho0/omega                * charged techni_rho                
361   f + fbar -> W_L+ W_L-        370   f + fbar' -> W_L+/- Z_L0       
362   f + fbar -> W_L+/- pi_T-/+   371   f + fbar' -> W_L+/- pi_T0      
363   f + fbar -> pi_T+ pi_T-      372   f + fbar' -> pi_T+/- Z_L0      
364   f + fbar -> gamma pi_T0      373   f + fbar' -> pi_T+/- pi_T0     
365   f + fbar -> gamma pi_T0'     374   f + fbar' -> gamma pi_T+/-     
366   f + fbar -> Z0 pi_T0         375   f + fbar' -> Z0 pi_T+/-        
367   f + fbar -> Z0 pi_T0'        376   f + fbar' -> W+/- pi_T0        
368   f + fbar -> W+/- pi_T-/+     377   f + fbar' -> W+/- pi_T0'       

\end{verbatim}
All of the processes from {\tt 361} to {\tt 377} can be accessed at once
by setting {\tt MSEL=50}.

The production and decay rates depend on several "Straw Man"
technicolor parameters ($D$ denotes the default value of
a parameter):

\begin{verbatim}
* Techniparticle masses
PMAS(51,1) (D=110.0 GeV)  neutral techni_pi mass
PMAS(52,1) (D=110.0 GeV)  charged techni_pi mass
PMAS(53,1) (D=110.0 GeV)  neutral techni_pi' mass
PMAS(54,1) (D=210.0 GeV)  neutral techni_rho mass
PMAS(55,1) (D=210.0 GeV)  charged techni_rho mass
PMAS(56,1) (D=210.0 GeV)  techni_omega mass

Note:  the rho and omega masses are not pole masses

* Lagrangian parameters
PARP(141) (D= 0.33333)  $\sin\chi$, the mixing angle between
   technipion interaction and mass eigenstates
PARP(142) (D=82.0000 GeV)   F_T, the technipion decay constant
PARP(143) (D= 1.3333)   Q_U, charge of up-type technifermion; the
   down-type technifermion has a charge Q_D=Q_U-1   
PARP(144) (D= 4.0000)   N_TC, number of technicolors
PARP(145) (D= 1.0000)   C_c, coefficient of the technipion decays to charm
PARP(146) (D= 1.0000)   C_b, coefficient of the technipion decays to bottom
PARP(147) (D= 0.0182)   C_t, coefficient of the technipion decays to top
PARP(148) (D= 1.0000)   C_tau, coefficient of the technipion decays to tau
PARP(149) (D=0.00000)   C_pi, coefficient of technipion decays to gg
PARP(150) (D=1.33333)   C_pi', coefficient of technipion' decays to gg
PARJ(172) (D=200.000 GeV) M_V, vector mass parameter for technivector
   decays to transverse gauge bosons and technipions
PARJ(173) (D=200.000 GeV) M_A, axial mass parameter
PARJ(174) (D=0.33300)   $\sin\chi'$, the mixing angle between
   the technipion' interaction and mass eigenstates 
PARJ(175) (D=0.05000)   isospin violating technirho/techniomega mixing 
   amplitude
\end{verbatim}

Note, the decays products of the $W$ and $Z$ bosons are distributed 
according to phase space, regardless of their designation as 
$W_L/Z_L$ or transverse gauge bosons.  The exact meaning of 
longitudinal or transverse polarizations in this case requires
more thought.

\end{document}